\pdfoutput=1

\documentclass[prl,twocolumn,showpacs,preprintnumbers,amsmath,amssymb,
superscriptaddress,nofootinbib]{revtex4-1}
\usepackage{epsfig}
\usepackage{graphicx}
\usepackage{color}
\usepackage{hyperref}

\usepackage{stackengine}
\stackMath
\usepackage{graphicx}

\begin{document}
\title{Scattering Amplitudes from Soft Theorems and Infrared Behavior}
\author{Laurentiu Rodina}
\affiliation{Institut de physique theorique, Universite Paris Saclay, CEA, CNRS, F-91191 Gif-sur-Yvette, France
}
\date{\today}
\begin{abstract} % abstract
We prove that soft theorems uniquely fix tree-level scattering amplitudes in a wide range of massless theories, including Yang-Mills, gravity, the non-linear sigma model, Dirac-Born-Infeld, dilaton effective theories, extended theories like NLSM$\oplus \phi^3$, as well as some higher derivative corrections to these theories. We conjecture the same is true even when imposing more general soft behavior, simply by assuming the existence of soft operators, or by imposing gauge invariance/the Adler zero only up to a finite order in soft expansions. Besides reproducing known amplitudes, this analysis reveals a new higher order correction to the NLSM, and two interesting facts: the subleading theorem for the dilaton, and the subsubleading theorem for DBI follow automatically from the more leading theorems. These results provide motivation that asymptotic symmetries contain enough information to fully fix a holographic S-matrix. 
\end{abstract}
\maketitle
\section{Motivation}
%%%%%%%%%%%

Three related concepts are central to quantum gravity: holography, the S-matrix, and the black hole information paradox. The holographic principle states that a theory with gravity in the bulk may be described completely by a non-gravitational quantum field theory on the boundary. This motivates how Hawking entropy can be proportional to the area of a black hole instead of its volume. 

The S-matrix on the other hand is the unique local gauge invariant observable of quantum gravity. The naive merger of quantum mechanics and general relativity leads to tensions with locality and unitarity, and the S-matrix should have a formulation which avoids these tensions. This is expected only because the S-matrix itself is a naturally holographic object: it describes the $\langle$in$|$out$\rangle$ matrix of states measured at asymptotic infinity. 

Recently, the holographic nature of the S-matrix was made even more concrete by demonstrating the equivalence between asymptotic symmetries and soft theorems \cite{Strominger2014,Strominger2013c,He2014a,He2014,Kapec2017}, opening new paths towards finding a holographic dual of flat space-time itself, some reviewed in \cite{Strominger2017}. Even more surprisingly, it was proposed that infrared considerations and asymptotic symmetries can have implications for the black hole information paradox \cite{Hawking2016b,Hawking2016c,Strominger2017a}. This raises an apparently superficial question: how much information can soft particles actually carry? 

The goal of this letter is to take inspiration from the above issues and ask a more well defined question, in the spirit of the S-matrix program: \\

{\em How much of an amplitude can be fixed by soft theorems?}\\

The naive answer is that the low energy (IR) behavior of amplitudes is completely disjoint from the high energy (UV) behavior, so soft particles can only carry some partial information, fixing only the IR part of an amplitude. The following separation seems valid then:
\begin{align}
\nonumber {A}=&\textrm{$A_\textrm{IR}$ (Soft theorem satisfying)} \\
&+ \textrm{$A_\textrm{UV}$ (Soft theorem avoiding)}\ .
\end{align}
But surprisingly, we find that the UV information is not inaccessible via soft theorems - it is simply hidden in several different soft limits. This implies that soft theorems are sufficient to fully fix scattering amplitudes! And in fact, even milder soft behavior can be used instead of the full soft theorems, and still we find the amplitudes are fixed. This enables us to discover scattering amplitudes starting purely from the assumption of soft operators, or what we will call ``soft gauge invariance" or ``soft Adler zero".

\section{Review of Soft Theorems}
Soft theorems describe a universal behavior of scattering amplitudes when the energies of one or more massless particles are taken to zero. This limit is taken by rescaling momenta with a soft parameter $p^\mu\rightarrow z p^\mu$, and taking the $z\rightarrow 0$ limit. The soft theorems then imply a factorization of the form:
\begin{align}
\label{full}
A_{n} \rightarrow \left(z^\sigma S^{(0)}+z^{\sigma+1}S^{(1)}+\ldots\right) A_{n-1}\ ,
\end{align}
where the $S_i$ are called soft operators, and encode symmetries of the theory being considered. 

Originally discovered for photons in \cite{Low1958} and extended to gravitons in \cite{Weinberg1965}, soft theorems have enjoyed a renewed interest, at least in part due to their uncovered equivalence to memory effects and asymptotic symmetries \cite{Strominger2016}, and the discovery of a new soft theorem for gravitons \cite{Cachazo2014b}. These results subsequently lead to an investigation of soft theorems and asymptotic symmetries for many other theories \cite{Casali2014,White2014,Campiglia2014,Campiglia2015,Campiglia2015a,Campiglia2016,Campiglia,Laddha2018,
Elvang2016c,Guerrieri2017a,Huang2016,Hamada2017,
Conde2017a,Mao2017,Mao2017a,Conde2017,
Lysov2015,Li2017a,
Pate2018,Pate2017,DiVecchia2015b,DiVecchia2015a,DiVecchia2016,DiVecchia2016a,DiVecchia2016b,DiVecchia2017a,
DiVecchia2017,Bianchi2014,Bianchi2016,
Schwab2014a,Chakrabarti2017a,Laddha2017a,Sen2017b,Sen2017a,Hamada2018a,Li2018a}. Soft theorems were shown to follow from considerations of gauge invariance, locality, unitarity, \cite{Broedel2014,Bern2014}, shift symmetries in the case of scalar theories \cite{Low2015,Low2017,Low2018}, the CHY formalism \cite{Cachazo2015a,Cachazo2014,Cachazo2015,Cachazo2013,Schwab2014,Volovich2015}, ambitwistors \cite{Lipstein2015,Geyer2014a}, or transmutation operators \cite{Cheung2017a}. Soft theorems for dilaton theories also hinted towards a hidden conformal symmetry in gravity \cite{Loebbert2018}. 

Meanwhile, a different type of soft behavior, known as the Adler zero \cite{Adler1965,Susskind1970}, was being exploited in the construction of various effective theories: the non-linear sigma model (NLSM), Dirac-Born-Infeld (DBI), the Galileon \cite{Dvali2000,Nicolis2008} and special Galileon (SG) (see \cite{Kampf2013a,Cachazo2015a} for overviews of these theories). This was done by constraining the theory space of possible effective theories  \cite{Cheung2015,Cheung2016,Cheung2014a,Cheung}, by direct construction using newly enabled recursions \cite{Cheung2016a,Luo2015}, and most recently through the ``soft bootstrap" procedure \cite{Elvang2018}. This special behavior was also used to rule out possible counterterms in $\mathcal{N}=8$ supergravity \cite{Arkani-Hamed2008}\cite{Chen2015}, recently shown to be finite up to 5 loops in $D=4$ \cite{Bern2018}.  

Soft limits as formal Taylor series were also crucial to proving that various scattering amplitudes can be fixed by only three conditions: ``weak" locality, gauge invariance/Adler zero, and minimal mass dimension \cite{Arkani-Hamed2016,Rodina2016a}. Similarly, in \cite{Rodina2016} it was conjectured that there exist unique objects satisfying locality and correct D-dimensional BCFW scaling \cite{Britto2005,Britto2004,Arkani-Hamed2008a}. Remarkably, in all of these cases unitarity emerged as a direct consequence of other basic principles. The present article is a continuation of this previous work, this time investigating the constraints imposed by locality and various types of soft behavior.

\section{Fixing amplitudes with soft behavior}
We start with a general local ansatz $B_n\equiv B_n\delta(\sum_i p_i)$, which has some appropriate singularity structure, corresponding to propagators of tree diagrams. Then we take $m$ particles soft, obtaining a soft limit expansion
\begin{align}
\label{exp}
B_n\rightarrow z^{\sigma} B_n^0+z^{\sigma+1} B_n^1+z^{\sigma+2} B_n^2+\ldots\ .
\end{align}
Finally we demand that this matches the corresponding soft theorems (which exist up to a finite order $N$)
\begin{align}
B_n\rightarrow 
\left(z^\sigma S_0+z^{\sigma+1}S_1+\ldots+z^{N}S_{N}\right)A_{n-m}+\ldots \ .
\end{align}
This leads to a system of equations which turns out to have a unique solution: $B_n=A_n$, the corresponding higher point amplitude.

But the constraints coming from IR behavior can be relaxed beyond what is dictated by soft theorems, and still we find that the UV part is fixed.  We can consider what sort of general local functions $B_n$ and $B_{n-m}$ can satisfy a soft behavior of the type
\begin{align}
B_{n}\rightarrow \left(z^\sigma S_0+z^{\sigma+1}S_1+\ldots+z^N S_{N}\right) B_{n-m}+\ldots \ .
\end{align}
This happens only if both $B_n$ and $B_{n-m}$ are scattering amplitudes! Since in this approach no previous knowledge of amplitudes is required, we can probe for new amplitudes, in particular higher order corrections, simply by assuming that such objects satisfy the same symmetries as the ``base" theory. 

Finally, we consider an even more general soft behavior, without any factorizing soft operators. In the soft expansion (\ref{exp}) we simply require $B_n^{0}$ up to $B_n^{N}$ to satisfy gauge invariance (for vector theories), or the Adler zero (for scalar theories) in the soft particles. Even this very weak condition is sufficient to fix $B_n=A_n$, but this time not including any higher order corrections. This is a direct improvement of the previous results in \cite{Arkani-Hamed2016,Rodina2016a}, since now we require gauge invariance/Adler zero only up to a finite order. 

In conclusion, we propose three new constraints pertaining purely to the IR behavior which, together with locality, completely fix various scattering amplitudes:
\begin{align*}
&\textrm{(1) Soft theorems}\\
&\textrm{(2) Soft operators}\\
&\textrm{(3) Soft gauge invariance/soft Adler zero.}
\end{align*}

The above results seem to hold for all massless theories which satisfy soft theorems. This includes QED, Yang-Mills, gravity, NLSM, DBI, dilaton effective theory \cite{Boels2015b,Bianchi2016,Huang2015,Elvang2013,Elvang2012}, among others. Most such amplitudes are of course already known, but we do discover a novel amplitude: the two extra derivative correction to the NLSM. Higher corrections to the NLSM were computed in \cite{Carrasco2016}, but those start at four extra derivatives. We are also able to reproduce the mysterious extended theories found in \cite{Cachazo2016}, simply by imposing the corresponding soft operators on an appropriate ansatz.

\section{Soft theorem avoiding terms and enhanced soft limits}
The argument for uniqueness from soft theorems is straightforward: assume there are two different objects, the amplitude $A_{n}$, and a general local function $B_{n}$, satisfying the same soft theorems, which exist up to some order $\mathcal{O}(z^N)$. This implies the difference $B_{n}-A_{n}$ must behave as $\mathcal{O}(z^{N+1})$ in all $n$ soft limits. Therefore, after imposing the soft theorems on $B_{n}$, it must take a form
\begin{align}
\label{key}
B_{n}\rightarrow \textrm{[piece satisfying soft theorems]}+[\mathcal{O}(z^{N+1})\textrm{ terms}] 
\end{align}
The question is then whether such $\mathcal{O}(z^{N+1})$ soft theorem avoiding terms can exist at the mass dimension being considered. If these terms do not exist, it follows that $B_{n}=A_{n}$, so there is a unique object satisfying the soft theorems.

Surprisingly, it turns out that for the most common amplitudes, and even their low lying derivative corrections, this is indeed the case.  Their mass dimension is too low to allow any soft theorem avoiding terms, so the amplitude is fully captured by its low energy behavior.

Proving this is slightly more complicated than simple power counting, because momentum conservation can lead to non-trivial cancellations in soft limits. It is well known in fact that various scalar amplitudes do enjoy enhanced soft limit behavior, different from what naive power counting suggests. In particular four such solutions are known: NLSM, DBI, the Galileon vertex, and the special Galileon. We will call such objects enjoying enhanced soft limits $A_{\textrm{enh}}$. 

Therefore, depending on the theory, multiplicity, and mass dimension, imposing the soft theorems can lead to one of three outcomes
\begin{align}
B_{n} =& A_{n}\ ,\\
B_{n}= & \textrm{IR}[A_{n}]+f(e,1/K) A^{\textrm{enh}}\ ,\\
\nonumber \textrm{or}\\
B_{n} =& \textrm{IR}[{A}_{n}]+\textrm{[trivial objects]}\ ,
\end{align}
where IR[$A_n$] is the IR piece of $A_n$ that is under the control of soft theorems, and $f(e,1/K)$ can be some function of polarization vectors and propagators, left-over after factoring out $A^\textrm{enh}$. By ``trivial objects" in the third equation we mean independent terms which by simple power counting avoid all imposed soft theorems. Which $A^\textrm{enh}$ could appear depends on the theory being considered. For example, because of the ordered propagator structure, starting from a Yang-Mills ansatz we can only get $A^{\textrm{NLSM}}$ as the enhanced object.

\section{Yang-Mills}
Yang-Mills amplitudes satisfy single soft theorems \cite{Casali2014}
\begin{align}
A_{n+1}\rightarrow \left( \frac{1}{z}S_0 +z^0S_1\right)A_n+\mathcal{O}(z)\ ,
\end{align}
which hold even for the higher derivative corrections. These correspond to higher mass dimension operators in the general effective Lagrangian \cite{Barreiro2013,Barreiro2012a,He2017}:
\begin{align}
L=F^2+a_0 F^3+a_1 F_1^4+a_2 F_2^4+a_3 F_3^4+a_4 F_4^4\ ,
\end{align}
where the $F_i^4$ operators represent the different possible contractions of field strengths.

Our goal is to obtain the various amplitudes by starting from a local general ansatz, $B_n(p^{n-2+\kappa})$, with cubic ordered propagators, and with $(n-2+\kappa)$ powers of momenta in the numerators. We introduce $\kappa$ to keep track of the extra number of derivatives in the operator considered: $\kappa=0$ corresponds to the usual YM amplitude; $\kappa=2$ corresponds to the amplitude with an $F^3$ operator insertion; for $\kappa=4$  there are five different amplitudes, corresponding to an $(F^3)^2$ or an $F^4$-type operator insertion.

\subsection{Soft theorems}
To illustrate the procedure, we first consider an $n=5$ example. Take a local function with $\kappa=0$, $B_5(p^3)$, and impose the soft theorems:
\begin{align}
B_5(p^3)&\underset{p_i\rightarrow 0}{\longrightarrow} \left(\frac{1}{z}S_0+z^0 S_1\right)A_4 \ ,
\end{align}
repeating the process for every particle. We find a unique solution satisfying these constraints, namely $B_5=A_5$, the $n=5$ Yang-Mills amplitude.

We can do the same starting with a higher mass dimension ansatz, $\kappa=2$, using the known amplitude with an $F^3$ operator insertion. Impose for each particle:
\begin{align}
B_5(p^5)&\longrightarrow \left(\frac{1}{z}S_0+z^0S_1\right)A_4^{F^3}\ ,
\end{align}
and the solution is again unique: $B_5=A_5^{F^3}$. 

We can go to an even higher mass dimension at $\kappa=4$, using the known lower point amplitudes $A^{\kappa=4}\equiv a_1 A^{(F^3)^2}+a_2 A^{F_1^4}+a_3A^{F_2^4}+a_4A^{F_3^4}+a_5A^{F_4^4}$, and imposing:
\begin{align}
\label{f4}
B_5(p^7)\longrightarrow \left(\frac{1}{z}S_0+z^0S_1\right)A_4^{\kappa=4}
\end{align} 
There are five solutions, as expected: $B_5=A_5^{\kappa=4}$.

Increasing the mass dimension to $\kappa=6$, we can finally encounter terms that trivially avoid the soft theorems, for example terms like:
\begin{align}
\label{extrivial}
 \frac{(e.e\, e.e\, e.p_5)( p_1.p_2\, p_3.p_4)^2}{p_1.p_2\, p_3.p_4}\ .
\end{align}

 In this case, the solution takes a less useful form:
\begin{align}
 B_5(p^9)= \textrm{IR}[A_5^{\kappa=6}]+\textrm{[trivial $\mathcal{O}(z)$ objects]}\ .
\end{align}

This behavior is true in general, except for two exceptions at low multiplicity. First, at $n=4$ we can only obtain the $\kappa=0$ case by imposing soft theorems. Higher values for $\kappa$ allow terms like $e_1.e_2 e_3.e_4 p_1.p_2$, which scale as $\mathcal{O}(z)$ in all soft limits.
The second exception is $n=6$, $\kappa=4$, where we find:
\begin{align}
B_6(p^{10})= \textrm{IR} [A_6^{F^4}]+f(e) A_6^{\textrm{NLSM}}\ ,
\end{align}
where $f(e)$ is a function only of the six polarization vectors. 

Except these two cases, we claim that soft theorems fully constrain amplitudes for $\kappa=0,2,\textrm{ and }4$, for all $n$. 

The proof follows immediately from counting arguments: at $n>6$, forming a trivially soft theorem avoiding term like (\ref{extrivial}) requires numerators of mass dimension $n+4$. This is because locality forces the existence of at least two 2-particles poles per term, which are singular in four different soft limits. The exception to this rule can only come from NLSM amplitudes, at even multiplicity, which require numerators of higher mass dimension  $n+(n-2)$, in order to cancel the extra propagators. Therefore, uniqueness from soft theorems follows as long as $n-2+\kappa<n+4$, so $\kappa<6$, for $n>6$.

\subsection{Soft operators}
We can replace the known lower point amplitudes with general local functions, and the process still works. Remarkably, this ends up fixing the lower point functions as well! 

Starting from general four and five point ansatze, impose:
\begin{align}
B_5(p^3)&\underset{p_1\rightarrow 0}{\longrightarrow} \left(\frac{1}{z}S_0+z^0S_1\right)B_4(2,3,4,5)\ ,
\end{align}
for each particle. The solution is $B_5=A_5$ as before, but we also recover $B_4=A_4$. 

This is still true for $\kappa=2$, where we find $B_5=A_5^{F^3}$, and also $B_4=A_4^{F^3}$. However, with $\kappa=4$, besides the usual amplitudes, we find an extra non-gauge invariant solution. It would be interesting to understand the origin of this object, and whether it persists at higher multiplicity.

As a result of these observations, we conjecture that at higher multiplicity soft operators fully fix the $\kappa=0, 2$ amplitudes, and $\kappa=4$ amplitudes up to possible extra non-physical solutions.

\subsection{Soft gauge invariance}
We can relax the IR behavior even further. Simply demanding that in the various soft expansions:
\begin{align}
B_5(p^3)&\underset{p_i\rightarrow 0}{\longrightarrow} \frac{1}{z}B_{5;i}^{-1}+z^0B_{5;i}^0+\ldots\ ,
\end{align}
the functions $B_{5;i}^{-1}$ and $B_{5;i}^0$ are gauge invariant in particle $i$, we find a unique solution $B_5=A_5$. We also conjecture that this remains true at higher multiplicity.

%%%%%%%%%%%%%%%%%%%%%%%
\section{Gravity}
%%%%%%%%%%%%%%%%%%%%%%%
Compared to Yang-Mills, gravity amplitudes satisfy one extra soft theorem \cite{Cachazo2014b}:
\begin{align}
A_{n+1}\rightarrow \left(\frac{1}{z} S_0 +z^0 S_1+z^1 S_2\right)A_n+\mathcal{O}(z^2)\ ,
\end{align}
but also have numerators of a higher mass dimension $(2n-4+\kappa)$. Again we use $\kappa$ to label different amplitudes, with $\kappa=0$ the usual GR amplitudes. The soft theorem avoiding terms in this case must scale as $\mathcal{O}(z^2)$ in all soft limits, corresponding to either DBI amplitudes or Galileon vertices as possible exceptions to power counting. 

Using the same counting arguments as for Yang-Mills, this implies that for $n=5$ and $n>6$, soft theorems fully constrain the $\kappa=0,2,\textrm{ and }4$ amplitudes. 

For $n=4$ only the usual $\kappa=0$ amplitude can be obtained via soft theorems, while at $n=6$ for $\kappa=4$ we find the solution:
\begin{align}
\label{extradbi}
B_6= \textrm{IR}[A_6^{\kappa=4}]+f(e) A_6^{\textrm{DBI}}\ .
\end{align}
Finally, we conjecture that the soft operator and soft gauge invariance constraints also work as they did in the Yang-Mills case. 

%%%%%%%%%%%%%%%%%%%%%
\section{NLSM}
%%%%%%%%%%%%%%%%%%%%%%%

Besides the single soft theorems known as the Adler zero \cite{Adler1965,Kampf2012,Kampf2013a}, the NLSM also satisfies double soft theorems:
\begin{align}
A_n\rightarrow \left(z^0S_0+z S_1\right)A_{n-2}+\mathcal{O}(z^2)\ ,
\end{align}
with explicit expressions for the soft factors given in \cite{Cachazo2015} for the adjacent limit, and \cite{Du2015} for the non-adjacent limit. In this case soft theorem avoiding terms must scale as  $\mathcal{O}(z^2)$ in {\em double} soft limits. 

The starting ansatz for the NLSM is also of the form $B_n(p^{n-2+\kappa})$, with numerators of mass dimension  $(n-2+\kappa)$, and poles corresponding to propagators of quartic diagrams.
Taking into account the minimum two three-particle poles per diagram, simple mass dimension counting proves that for $n\ge 8$ soft theorems must fully constrain the $\kappa=0,2, \textrm{ and }4$ cases. It should be noted that $\kappa=4$ only follows once we use the non-adjacent soft theorem. For $n=6$ only the $\kappa=0, 2$ cases can be obtained.

We can check that this reproduces the higher corrections to the NLSM computed in \cite{Carrasco2016}, but those start out at $\kappa=4$, so we have nothing check against for $\kappa=2$. However, we can use the soft operator approach to find if any $\kappa=2$ amplitudes exist.

\subsection{Soft operators}
As before, we conjecture that soft operators reproduce the known $\kappa=0$ and $\kappa=4$ solutions (for $n>6$).  But more interestingly they can also produce  the $\kappa=2$ corrections that we were missing.  By simply imposing
\begin{align}
B_6^{\kappa=2}\rightarrow \left(z^0S_0+z S_1\right)B_4^{\kappa=2}\ ,
\end{align}
we obtain unique solutions, with the four point amplitude given by the cyclically invariant combination:
\begin{align}
\label{newnlsm}
 A_4^{\kappa=2}=&s_{12}s_{14}\ .
\end{align}
It is interesting to note that $Z$-theory produces higher order corrections all of which obey BCJ relations \cite{Bern2008,Chen2013,Du2016,Carrasco2016,Carrasco2016a}, while the solution corresponding to (\ref{newnlsm}) does not.

\subsection{Soft Adler zero}
Transferring our insights from the previous examples, it turns out that we can also impose a ``soft Adler zero" to obtain the NLSM amplitude. Taking a formal double limit in particles $n$ and $n-1$:
\begin{align}
B_{n}\rightarrow \frac{1}{z} B_{n-2}^{-1}+z^0 B_{n-2}^0+z B_{n-2}^1\ ,
\end{align}
we now impose that the three terms above have $\mathcal{O}(z)$ behavior when taking $n$ and $n-1$ soft. Repeating the procedure for the other particles, we obtain that $B_n$ must be the NLSM amplitude.

\subsection{Single soft theorems and extended NLSM}
In \cite{Cachazo2016} it was discovered that various amplitudes contain hidden so called extended theories in their single soft limits. In the case of the NLSM, schematically this limit is:
\begin{align}
\label{ext1}
A_{n}=z \sum_i  s_{in}A_{n-1}^{\textrm{NLSM} \oplus\phi^3}(i)+\mathcal{O}(z^2)\ ,
\end{align}
where $A_n^{\textrm{NLSM}\oplus \phi}$ is an amplitude of pions interacting with a scalar bi-adjoint theory. Its Feynman rules were found in \cite{Low2017}, but using our guiding principle of soft behavior we can derive even these mixed amplitudes simply by assuming that each NLSM scalar obeys the NLSM soft theorems/operators.

%%%%%%%%%%%%%%%%%%%%%%%%%%%%%
\section{Dirac-Born-Infeld}
%%%%%%%%%%%%%%%%%%%%%%%%%%%%%
DBI satisfies double soft theorems up to order $\mathcal{O}(z^3)$ \cite{Cachazo2015}, so we are looking for the lowest mass dimension objects with  $\mathcal{O}(z^4)$ in all {\em double} soft limits. This is again the Gallileon vertex,  which has a mass dimension $(2n-2)$. 

The DBI ansatz has a form $B_n(p^{2n-4+\kappa})$ with quartic propagators, so taking into account the minimum two three-particle poles for $n\ge 8$, this implies we can obtain the $\kappa=0,2, \textrm{ and }4$ cases from soft theorems before hitting Galileon vertices. We conjecture the same cases can be obtained from soft operators (with the usual caveats), and that the soft Adler zero implies the $\kappa=0$ case.

Because DBI obeys soft theorems up to such a high order, an interesting stronger statement holds: $\kappa=0$ DBI is completely fixed by just the leading and subleading theorems, implying that the subsubleading theorem is in fact not independent. As will be discussed next, a similar fact is true for the dilaton theories.

%%%%%%%%%%%%%%%%%%%%%%%%%%%%%
\section{Dilaton}
%%%%%%%%%%%%%%%%%%%%%%%%%%%%%
Due to its peculiar universality, the dilaton is also interesting to study from the perspective of its soft behavior. First, we should note there are two types of dilatons: a gravity dilaton \cite{DiVecchia2017a,DiVecchia2016,DiVecchia2015a,DiVecchia2016a}, and a dilaton associated to the  breaking of conformal invariance \cite{DiVecchia2017,Elvang2013,Elvang2012}. We will focus on the latter, with soft theorems:
\begin{align}
A_n\rightarrow \left(z^0S_0+z^1S_1\right)A_{n-1}\ ,
\end{align}
explicitly given in  \cite{Bianchi2016}\cite{Huang2015}.

Depending on how conformal invariance is broken, two classes of theories emerge, both of which we can recover. In the case of spontaneous breaking, the dilaton is a Goldstone boson in the spectrum of the theory, whereas in the case of explicit breaking it can be thought of as a external source. Therefore we can control which case we consider by the form of our ansatz: pure polynomials will reproduce explicit breaking, whereas allowing propagators reproduces the spontaneous case. 

Using the previous reasoning, uniqueness from soft theorems (operators) follows if there are no terms with $\mathcal{O}(z^2)$ behavior in all single soft limits. For explicit breaking this is the Galileon vertex, while for spontaneous breaking it is the DBI amplitude.

\subsubsection{Explicit breaking}
In this case we consider a polynomial ansatz $P_n(p^{4+\kappa})$. The mass dimension of the Galileon is $(2n-2)$, so for $4+\kappa<2n-2$ all amplitudes are fixed by soft theorems. This implies that we can obtain higher and higher corrections as we increase the multiplicity. On the other hand, this also implies that the subleading theorem is not independent. The leading order theorem is sufficient as long as $4+\kappa<n$, so for all of these cases scale invariance implies conformal invariance. Since we do not assume unitarity, this generalizes the result of \cite{Bianchi2016}, where this observation was first made.

The soft operator approach also reproduces the known results. By simply imposing:
\begin{align}
\label{softp}
P_n\rightarrow \left(z^0 S_0+zS_1\right)P_{n-1}\ ,
\end{align}
we have checked that we reproduce the results in \cite{Elvang2013,Elvang2012}, for $n>5$, $\kappa=0,\ 2$.

For higher $\kappa$, we conjecture that soft operators fix amplitudes within the same bound as the soft theorems, that is for $4+\kappa<2n-2$.

\subsubsection{Spontaneous breaking} 
For this case we also allow three-particle poles in an ansatz $B_n(p^{4+\kappa})$, and impose the soft theorems. For $\kappa=0$ the solutions in the spontaneous case are identical to the explicit case. For $\kappa=2$ at $n=6$, we obtain the general solution:
\begin{align}
B_6(p^6)=b_1A_6^\textrm{explicit}(p^6)+ b_2A_6^{\textrm{DBI}}\ ,
\end{align}
which is as expected, since the DBI action appears manifestly in the dilaton effective action \cite{Huang2015}.

\section{Conclusions}
We have shown that imposing soft theorems completely fixes a wide range of scattering amplitudes. Further, we have conjectured that in some cases the same should hold when imposing the less constraining soft operators and soft gauge invariance/Adler zero.

For both conceptual and practical purposes, it would be extremely useful to transform these uniqueness results into a general Inverse Soft Limit type construction \cite{Arkani-Hamed2009,Nguyen2009,Boucher-Veronneau2011,Nandan2012}, which would allow building amplitudes directly from soft factors. This in turn could  make manifest yet a new facet of scattering amplitudes, that of asymptotic symmetries, perhaps leading to a purely holographic description of the S-matrix.

\section{Acknowledgements}
The author would like to thank Nima Arkani-Hamed and Song He for discussions, comments, and collaboration on related work; John Joseph Carrasco for discussions, and Congkao Wen for discussions and comments on the manuscript. The author is grateful to the Institute of Theoretical Physics, Chinese Academy of Sciences, Beijing, where part of this work was completed. This work is supported by the European Research Council under ERC-STG-639729, ``preQFT: Strategic Predictions for Quantum Field Theories".

\bibliographystyle{apsrev4-1}

\bibliography{softprl}

\end{document}